# A Longitudinal Study of Identifying and Paying Down Architectural Debt

Maleknaz Nayebi
Ecole Polytechnique de Montreal
Email: mnayebi@polymtl.ca

Yuanfang Cai
Drexel University
Email: yfcai@cs.drexel.edu

Rick Kazman
University of Hawaii
Email: kazman@hawaii.edu

Guenther Ruhe
University of Calgary
Email: ruhe@ucalgary.ca

Qiong Feng
Drexel University
Email: qf28@cs.drexel.edu

Chris Carlson
Brightsquid
chris.carlson@brightsquid.com

Francis Chew
Brightsquid
francis@brightsquid.com

*Abstract*—Architectural debt is a form of technical debt that derives from the gap between the architectural design of the system as it "should be" compared to "as it is". We measured architecture debt in two ways: 1) in terms of system-wide coupling measures, and 2) in terms of the number and severity of architectural flaws. In recent work it was shown that the amount of architectural debt has a huge impact on software maintainability and evolution. Consequently, detecting and reducing the debt is expected to make software more amenable to change. This paper reports on a longitudinal study of a healthcare communications product created by Brightsquid Secure Communications Corp. This start-up company is facing the typical trade-off problem of desiring responsiveness to change requests, but wanting to avoid the ever-increasing effort that the accumulation of quick-and-dirty changes eventually incurs. In the first stage of the study, we analyzed the status of the "before" system, which indicated the impacts of change requests. This initial study motivated a more in-depth analysis of architectural debt. The results of this analysis were used to motivate a comprehensive refactoring of the software system. The third phase of the study was a follow-on architectural debt analysis which quantified the improvements made. Using this quantitative evidence, augmented by qualitative evidence gathered from in-depth interviews with Brightsquid's architects, we present lessons learned about the costs and benefits of paying down architecture debt in practice.

*Index Terms*—Architectural debt, Cost-benefit analysis, Longitudinal study, Refactoring

## I. Introduction

Recent research [9], [25] has shown that architectural design flaws accumulate in software projects over time, and that the accumulation of these flaws creates a specific kind of technical debt [2] that we call *architectural debt*. Architectural debt exists and grows because design flaws are easy to introduce unnoticed; they are introduced by the maintenance activities of programmers as they go about their "main" business of adding features and fixing bugs. These design flaws erode the quality of a software system and propagate bugginess among the system's source files. These flawed structures have been shown to incur high maintenance penalties [6], [16]—increased numbers of bugs, increased numbers of changes, and consequently more lines of code committed and more effort. This additional effort is the interest that a project pays on the incurred debt. Removing these flaws requires effort, in the form of refactoring, and the benefits of refactoring have historically been difficult for architects and project managers to quantify or justify; they simply do not have the data and analytic tools available to quantify the costs and benefits of a proposed refactoring. For this reason, large-scale refactorings to remove debt are exceedingly rare.

In this paper we report on the results of a longitudinal study at a company that produces secure communication software for health information. Brightsquid[1] is a global provider of HIPAA-compliant[2] communication solutions, providing compliant messaging and large file transfer for medical and dental professionals since 2009. Secure-Mail is Brightsquid's core communication and collaboration platform. It offers role-based API access to a catalog of services and automated workflows. Their platform supports aggregating, generating, and sharing protected health information across communities of health care patients, practitioners, and organizations. Brightsquid has been working on a number of projects to achieve these business goals, and this study is focused on analyzing their core software platform. The company is facing the typical problem of software start-ups: they need to quickly enter a competitive market with innovative product ideas to produce revenue in the near-term; simply put they need to satisfy current users and their expectations. At the same time, the company is facing the demands of growing their customer base and satisfying their requirements [8]. As Brightsquid's product manager puts it: "The job of the start-up is to find a sustainable business model—in other words, to discover an important and urgent problem, that a defined and accessible segment of customers will pay for to have solved. The likelihood of finding the right problem and customer segment, let alone building the right and enduring technology solution on the first try is about 0%. This means that if ignored, the likelihood of architectural debt in a start-up is conversely 100%. Many start-ups embrace agile software development methodologies, where the typical attitude is that solution architecture evolves organically. Evolution is at the mercy a continually changing environment, so key start-up characteristics comprise survival, flexibility, speed, and a revolving door of opportunities, stakeholders and employees."

---

[1] https://Brightsquid.com/
[2] HIPPA: Health Insurance Portability and Accountability Act of 1996

The paper reports on the results of a longitudinal study of Brightsquid's main software platform. In this study we performed an architectural analysis of this platform before and after refactoring, with the goal of identifying and quantifying the architectural debt in the before and after states. To the best of our knowledge, this is the first real-world empirical study of architectural debt over a long period of time with the goal of demonstrating the benefit of improving a product's software architecture by paying down architectural debt (through refactoring).

Specifically this study is focused on answering three main research questions:

**RQ1:** do quantitative measures of architecture complexity change significantly before/after refactoring?

**RQ2:** do quantitative project quality measures change significantly before/after refactoring?

**RQ3:** do qualitative perceptions of architectural quality change before/after and does this match the quantitative changes? The results of this longitudinal study are, we believe, quite dramatic. Brightsquid, by paying down its architectural debt, improved the maintainability of their code based significantly. Velocity went up significantly: the average time to resolve new issues in the after version went down by 72% and build time was reduced by over 83%, as compared with the before version. In addition, the number of bugs resolved per month nearly doubled and the lines of code required to make these fixes were reduced by 2/3.

In the remainder of this paper we will describe Brightsquid's business context, the details of the longitudinal analyses that we conducted, and the results that we obtained.

## II. Context and Baseline Analysis

This study has been done as part of a three year collaborative program supported by the Canadian Natural Sciences and Engineering Research Council (NSERC) designed to analyze the impact of code changes at Brightsquid. The project had multiple phases and was kicked off in the Summer of 2016. As part of this project, we studied the status of architectural debt and its impact on code maintainability at Brightsquid [28]. The timeline of this project in regards to the scope of the paper is presented in Figure 1.

Inspired by the work of Begel and Zimmermann [1], in the initial phase of this project we performed a survey that included all of Brightsquid's developers, project and product managers (a total of nine employees) to pinpoint the most interesting questions in the domain of the project. Among the 21 stated questions the most frequently asked questions were:

*"How extensible is Brightsquid's software in comparison to some type of recognized standard?"*

*"What is the general cost of change on this software in comparison with some type of accepted standard?"*

*"What areas of code / services are non-performant?"*

*"What areas of the code-base are not utilized?"*

To answer these questions, we performed a preliminary analysis of the code for the "Platform" project which includes the main shared functionality of the project. We analyzed the code changes maintained in GITHUB and traced the changes to the change requests maintained on the project's JIRA issue tracking system, looking into all the files and file types (`Java` and `Javascript`). The overview of the results is shown in Figure 2. We found that the 10% of the commits with the highest churn (changed lines of code) were applied on just 270 files (2.5% of all files). The 25% of the commits with the highest churn were applied on 26.1% of the files (2,870 files). Our results also showed that 27.1% of the files (2,977 files) have never been changed after creation. And we found that 0.4% of all the files have changed with all the change requests. The distribution of the churn for all the files is shown in Figure 1 - (a). As we will show in Section V, these are all symptoms of architecture debt. For example, these 0.4% of files that are constantly changing are all members of architecture roots [23].

Among all the files in the Platform project, 49.1% of the files have only a single contributor and were never touched by any other person than the file creator. Looking across all releases and all the files that have been changed by each pair of non-consecutive releases showed that there is a 34% overlap between changed files on average. If we include also consecutive releases, this number goes up to 41%. In addition, we present a heatmap for the mutual file changes in Figure 2 - (b). Figure 2 - (c) shows the types of issues with respect to code churn for the $10^{th}$ percentile of the churn.

These initial analyses led us to believe that Brightsquid's code base had serious structural problems, which motivated us to perform an architectural analysis.

## III. Architectural Debt Analysis

In our architectural analysis, we attempted to detect, measure, and assess the consequences of architecture debt in two ways: 1) by calculating architecture-based maintainability metrics on the before and after versions of Brightsquid's software, and 2) by identifying the architectural flaws and architecture roots in their software.

### A. Maintainability metrics suite

We employed two state-of-the-art maintainability metrics to measure and compare architecture maintainability before and after refactoring:

*Decoupling Level (DL)*, introduced by Mo et al. [17]. Decoupling Level measures how well a software system is decoupled into independent modules, using Baldwin and Clark's design rule theory as the underlying theoretical foundation: the more active, independent, and small modules there are, the higher option values can be produced. Based on this rationale, Mo et al.'s algorithm first clusters source files into a design rule hierarchy (DRH), a hierarchical structure in which (1) files in lower layers can only depend on files in higher layers; (2) files within the same layer are clustered into mutually independent modules. Based on DRH, DL is calculated based on the design rule theory: the larger a module is, the lower its DL; the more independent modules there are, the higher the DL; the more dependents a module has, the lower its DL.

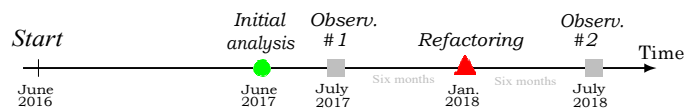

Fig. 1: Time line of the collaborative project

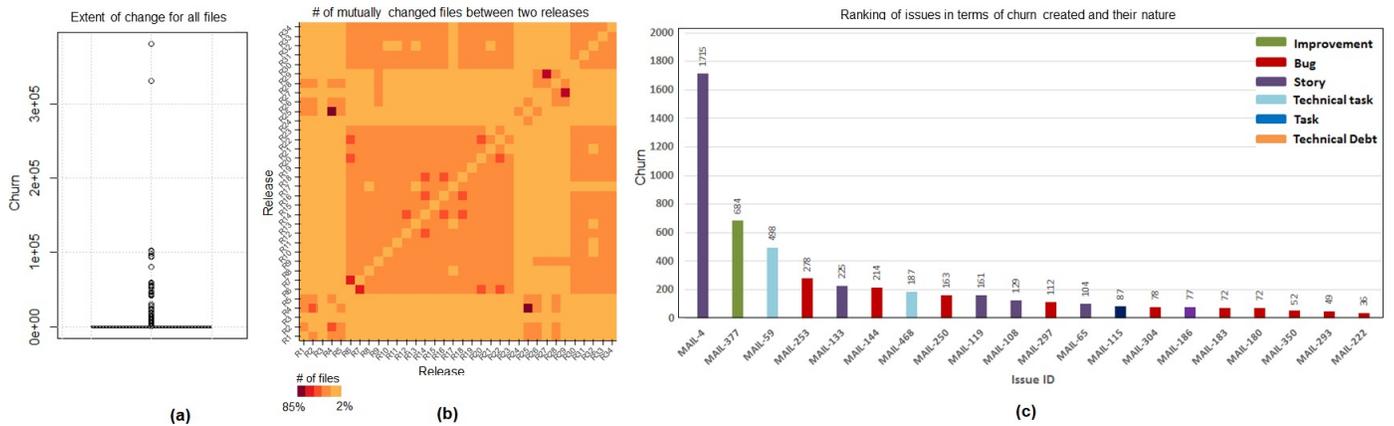

Fig. 2: Results of preliminary study at Brightsquid (a) churn distribution of all the files (b) Heatmap of co-changed files between releases (c) Type of change requests for top 10% of file with highest churn.

*Propagation Cost (PC)*, proposed by MacCormack et al [14] to measure how tightly coupled a system is. The calculation of PC is based on a matrix model of the dependencies among files. The rows and columns of the matrix are labeled with the files in the same order, and each nonempty cell in the matrix indicates an indirect or direct dependency between the file on the row and the file on the column. MacCormack et al.'s algorithm starts from the direct dependency relations among files in a system, and calculates the transitive closure of the matrix by add indirect dependencies to it until no more dependencies can be added. The final matrix thus contains all direct and indirect dependencies, PC is calculated as the number of nonempty cells divided by the total number of cells in the final matrix. PC has been used by researchers and practitioners to analyze large projects with similar domains and sizes [13].

These two metrics measure software from two complementary aspects: the level of decoupling vs. the level of coupling. In all systems the higher the DL, and lower the PC, and vice versa.

### B. Architectural Flaws

Following the work of Mo et al. [16], we detected the following 6 types of architectural design flaws from Brighsquid's software base:

**Clique:** A group files that are interconnected, forming a strongly connected component.

**Package cycle:** typically the package structure of a software system should form a hierarchical structure. A cycle among packages is therefore considered to be harmful.

**Improper inheritance:** we consider an inheritance hierarchy to be problematic if it falls into one of the following cases:
(1) a parent class depends on one or more of its children;
(2) the client of the class hierarchy uses/calls both a parent and one or more of its children, thus violating the Liskov Substitution Principle.

**Modularity violation:** properly designed modules designed with information hiding in mind should be able to change independently from each other. If two structurally independent modules in a DRSpace are shown to change together frequently in the revision history, it means that they are not truly independent from each other. We observe that in many of these cases, these modules have harmful implicit dependencies that should be removed. We call this flaw modularity violation. In this project, since the number of changes and co-changes are few, due to the relatively short revision history, we consider two files to have modularity violations if they have changed together at least 2 times but have no structural dependency on each other.

**Crossing:** if a file has many dependents and depends on many other files, that is, having high fan-in and high fan-out, then this file will appear to be at the center of a cross in its DSM. If the file at the center also changes frequently with it dependents and the files it depends on, we call these files form a Crossing. An example crossing from Brightsquid's platfom is shown in Figure 3.

**Unstable interface:** if a highly influential file is changed frequently with other files that directly or indirectly depend on it, then we call it an Unstable Interface. In this project, we consider a file to be an unstable interface if it changes together with at least 5 other files two times or more.

### C. Architecture Roots

Xiao et al. [23] proposed a software architecture model—*design rule spaces* (DRSpaces)—where each DRSpace models one aspect of a system, such as a feature, a pattern, etc. Using DRSpaces as an analytic lens, a software architecture can be viewed and analyzed as multiple overlapping DRSpaces. They also defined a new concept called *architecture roots*(or *roots* for short). These roots are the DRSpaces capturing the most error-prone files in a project. Dozens of studies have confirmed that five roots can typically cover 50% to 90% of the most error-prone files in a system. This implies that most error-prone files are usually architecturally connected; the more error-prone files are, the more likely that these files are architecturally connected so that bugs propagate through the connections. Following recent work [10], [18], we also detected the set of roots that cover at least 80% of all the error-prone files, before and after refactoring. By comparing the detected roots in the before and after versions of Brightsquid's software, as well as by interviewing the practitioners, we could determine if these roots could pinpoint areas of the system most in need of refactoring.

## IV. Organization of the Study

### A. Main Phases

The research study commenced in June 2016. Its main objective was to provide a methodology that we called *Intelligent Change Management* that would aid Brightsquid in responding more quickly to requested changes. The cornerstone of this strategy was to identify and pay down architectural debt so that the company could increase its feature-delivery and bug-fixing velocity.

From the perspective of identifying and paying down architectural debt, and referring to Figure 1, the study was structured into three phases:

1) **Baseline analysis** (June 2016 to May 2017.) During this time the data and key findings from Phase 1, as described in Section II, were collected and analyzed. These findings served as a strong motivation for the company to look more deeply into technical debt and its root causes and a decision was made to analyze the root causes of architecture debt.

2) **Architectural analysis before refactoring.** (July 2017.) At this time an in-depth architectural analysis was performed. The key attributes studied in this architectural debt analysis were those described in Section III: we captured DL and PC scores, we calculated the architecture roots, and we calculated the architecture flaws.

3) **Refactoring and repeated architectural debt analysis.** (January to August 2018.) During this period, given the results from Phase 2, an extensive architectural refactoring was undertaken by Brightsquid. The refactoring done in Phase 3 served multiple purposes: (i) reducing the technical (architectural) debt that had accumulated over time, (ii) adding new functionality in response to major emergent customer requirements, and (iii) fixing bugs. This process included purging of packages, consolidating tightly coupled functionality together, cleaning up inheritance structures, and purging complex and obsolete business logic that had accumulated over the years.

### B. The Process of Architectural Debt Analysis

To assess if and how the refactoring activity had affected the architecture, and, most importantly, the maintainability of the architecture, we conducted both quantitative and qualitative analyses. For the qualitative analysis, which is independent of the architectural debt analysis, we conducted a survey and multiple interviews to collect practitioner opinion and experience so that we could better understand the implications and outcomes of the refactoring activity. For example, we wanted to learn if the Brightsquid developers feel that it has become easier to maintain the system after refactoring, and if they thought that the 3 months of refactoring was worthwhile.

Fig. 3: A Design Structure Matrix with a Crossing Flaw (highlighted)

For quantitative analysis, we analyzed two versions of the system—before and after refactoring—as well as 6 months of revision history after each of these versions. In both analyses, we employed DV8[3]—a commercial version of the Titan architectural analysis tool suite [23], [24]—to analyze the architecture of each system version to assess architecture debt. To support this analysis we collected project history data, including issue records and git logs, so that we could quantitatively measure how maintenance activities have changed before and after refactoring.

Since the refactoring began on January 8th, 2018, for the before-refactoring analysis we analyzed the evolution history of the system 6 months prior to January 8th, and used a release in July 2017 as the target subject for architecture analysis of the "before" state. The refactoring was completed by March 1st, 2018. We analyzed the version released on that date, and 6 months of project history after that date, to assess the impacts of the refactoring.

For each of these two snapshots, we used DV8 to analyze the architecture from the three aspects elaborated in the previous section, collecting: 1) DL and PC scores, 2) the number of instances of architecture flaws and their scopes, and 3) the instances of architectural roots and their scopes. In particular, we were keen to know if the architectural problems we identified in the "before" state had been resolved during the refactoring process and if new problems emerged after refactoring. The output of this analysis will allow us to answer RQ1.

For these two periods of history, in addition to counting the total numbers of issues opened and fixed and the numbers of bug issues opened and fixed, we also calculated the LOC spent to fix each bug. We also counted the average numbers of days required for bugs to be fixed in the before and after versions. Our rationale for these measures is that if the architectural refactoring was successful it would become easier for developers to find and fix bugs and to develop new features. In this case the time and LOC spent should be significantly shortened after refactoring. The output of this analysis will allow us to answer RQ2.

## V. ANSWERS TO RESEARCH QUESTIONS

Now we present our results, organized according to the research questions stated in Section I. The first research question aims to quantitatively measure changes in architectural debts as reflected in Brightsquid's source code; the second question aims to quantitatively measure changes in maintainability and productivity outcomes as reflected in revision history; and the third research question explores the experiences from the developers and assess if the objective numbers match developers' experiences and intuitions.

[3]http://www.archdia.net

```
RQ1: do quantitative measures of
architecture debt change significantly
before/after refactoring?
```

We summarize architecture measures and debts in Table I and Table II.

After refactoring, the size of the code files shrunk by 41.5% while we found only three roots (containing 296 Files) that collectively account for 80% of bug fixes. By contrast there were five roots (including 295 files) before refactoring. In total, 37% of the files covered 80% of the bugs after refactoring while the number was 17% before the refactoring. After comparing these roots, we realize that some of the roots remain in the after version, meaning that the focal points of the system are centered around these 296 files, which could be determined by the nature of the application. Here we observe that the Pareto rule applies for architectural debt: the top few architectural roots always count for about 80% of the bugs, either before or after refactoring.

Analysis of the decoupling level showed that the modularity of the system decreased slightly, as the DL score reduced by 3%, and the PC score remained the same, at 6%. The small difference of DL could be caused by the fact that many redundant components (which may be independent, and contributed to a higher DL) were removed during the refactoring. But a difference of 3% is essentially noise.

These system-wide measures—the DL/PC scores, as well as the root analysis—do not reflect the changes in architecture directly. These are overall *average* health measures. But in architectural health the architectural flaws, which provide a more fine-grained analysis of architecture, changed drastically. To make an analogy, a human might be *mostly* healthy—having good blood pressure, low cholesterol, proper kidney function, etc.—but a brain tumor can undermine and render irrelevant all of those other measures.

TABLE I: Architectural analysis before and after refactoring.

| General information | Before | After |
| --- | --- | --- |
| # of files | 1713 | 711 |
| # of roots covering 80% of bugs | 5 | 3 |
| # of files in roots covering 80% of bugs | 296 | 295 |
| # of files covering 80% of bugs | 17% | 37% |
| **Architectural Metrics** | **Before** | **After** |
| Decoupling level | 86% | 83% |
| Propagation cost | 6% | 6% |
| **Architectural flaws** | **Before** | **After** |
| # of cliques | 17 | 10 |
| # of files influenced by cliques | 71 | 26 |
| # of unhealthy inheritance | 60 | 30 |
| # of files influenced by unhealthy inheritance | 222 | 102 |
| # of unstable interface | 12 | 8 |
| # of files influenced by unstable interface | 471 | 59 |
| # of crossings | 29 | 6 |
| # of files influenced by crossings | 387 | 47 |
| # of package cycles | 34 | 19 |
| # of files influenced by package cycles | 242 | 94 |

Concretely, the number of cliques was reduced by 41.1% as the # of files impacted by them were reduced by 63.3%. Both the number of unhealthy inheritances as well as the number of files affected by them reduced by almost 50%. The number of unstable interfaces was reduced from 12 to 8, and the number of files influenced by these interfaces reduced by 54%. The 79.3% reduction in the number of crossings was accompanied by a 87.8% reduction in the number of files impacted by the crossings. Finally, the number of package cycles reduced by 44.1% which shrunk the number of impacted files by 61.1%. Since the file names changed drastically before and after refactoring, it is impossible to compare the changes in terms of modularity violation instances. The comparative results are shown in Table I.

Why should we care about reducing flaws? Let us consider a specific example: the crossing shown in Figure 3. This crossing is centered around the file "User.java". In the before state of the system User.java was the center of a crossing containing 107 files. That is, in the before state, 106 other files either depended on User.java, or User.java depended on them. If User.java changed frequently—and it did—then many other files were potentially affected. In the after state, just 35 other files are coupled with User.java. This kind of reduction in complexity that we achieve by paying down the debt associated with architectural flaws means that, in general, changes are less likely to "ripple" to other files which in turn reduces the cost and complexity of those changes.

The refactoring activities were recorded as 106 change requests, which resulted in an effort of 563.8 person hours. We linked these refactoring issues to the commits that were related to 7 cliques, 23 crossings, and 4 unstable interfaces that we detected from architectural debt analysis before refactoring. We found that about 34% of the commits, 28% of the time, and 37% of lines of code during the refactoring period were related to the removal of architectural debt.

We looked into the time to fix issues in the files related to the refactored cliques, crossings, and unstable interfaces. We could find 13 issues after refactoring and compared them with the 51 issues related to these files before refactoring. Our comparison showed that the average time to close the issues relate to these critical files dropped by 72%.

> The average time needed to close issues before and after refactoring was reduced by 72%.

`RQ2: do quantitative project quality measures change significantly before/after refactoring?`

If the refactoring was successful, it should become *easier* for developers to add features or fix bugs. Here we use two proxies to quantify the *ease* of performing maintenance activities:

(1) # churn (lines of code changed) per issue. The rationale here is that the easier it is to fix a bug or to add new features, the fewer LOC will be needed to close a change request. If we only consider bug issues: the average bug-fixing churn after refactoring is about 34 LOC per issue on average, compared to 102 LOC before refactoring. It appears, and the development team believed, that the improved architecture made bug-fixing substantially more efficient. If we consider all issue types, and not just bugs, the averages are still improved: 208 LOC before refactoring and 156 after refactoring.

(2) # days needed per issue. The rationale here is that the easier it is to fix a bug or to add new features, the less time needed to close a change request. The data shows that the average bug-fixing duration reduced 30%, dropping from 10 days before to 7 days after. The productivity of the team improved from both reduced build time and reduced bug-fixing duration. The box plot distribution of the time spent to close a change request before and after refactoring and the churn are shown in Figure 4.

We also observed that before refactoring, 71 change requests (including 24 bug reports), involving code changes in the platform were resolved in five months, compared to 150 change requests (including 78 bug issues) after refactoring in a similar five month period. These numbers indicate that—with the same size team—after refactoring more issues are addressed and less code is "*spent*" to address each issue.

> The time and lines of code needed to close change requests are significantly lower than the required time before refactoring (p-value = 0.001 and 0.002). Considering these measures as proxies for productivity, it appears that the analysis of architectural debt correctly directed the architectural refactoring to increase developer productivity.

So far these quantitative analyses indicate a very successful refactoring activity, both reflected in the record of revision history and in the significant reduction of architectural flaws. Next we assess if the numbers match the developers' intuitions.

`RQ3: do qualitative perceptions of architectural quality change before/after and does this match the quantitative changes?`

To collect feedback from the development team, we conducted a structured interview. The goal of this interview was to qualitatively assess the *perceived* impact of refactoring. We wanted to understand if maintaining the system became noticeably better, in terms of bug-fixing and productivity. And we wanted to understand how much the developers thought that these improvements were triggered by the results of the Phase 2 analysis. To do this we interviewed key project stakeholders. The product manager, architect, and two back-end developers participated in the interview. Below, we first discuss the structured interview results, along with the descriptions provided by the participants, and then summarize the results of the follow-on unstructured interview.

TABLE II: Maintainability measures of Brightsquid's projects before and after refactoring.

| Measure | Before | After |
|---|---|---|
| # of files | 1713 | 711 |
| # of issues opened | 680 | 843 |
| # of issues fixed | 583 | 653 |
| # of bugs opened | 157 | 310 |
| # of bugs fixed | 137 | 267 |
| # of bugs that changed code in platform files | 24 | 78 |
| Amount of churn per bug | 102 | 33.9 |
| Average bug fixing time | 10.74 | 7.31 |

All participants were asked seven yes-no questions, where they were also given the opportunity to add some elaborations on their reasons for the answers given. The product manager, software architect, and two back-end developers attended this interview. We asked them to consider all the questions in terms of the system state *six months* before and *six months* after refactoring.

While our quantitative results showed that "*the # of change request opened before refactoring is less than the # of change requests opened after refactoring*" in Brightsquid, the product manager and a developer considered this a misleading characterization. The product manager argued that intuitively this seemed wrong as after refactoring the architecture and business rules were simplified and having less code, tables, and tests should result in fewer issues. The architect attributed the main cause of this increase to the introduction of new features. When comparing the status of change requests before and after refactoring, all the participants agreed with our finding that "*the portion of fixed change requests is higher after refactoring.*" The implication is that a desirable outcome had been achieved: more features were able to be added (with the same team size) after refactoring.

One of the surprising results was that "*the number of bugs opened after refactoring is significantly more than the number of bugs before.*" The software architect considered this to be consistent with his own perception. He stated that adding a large amount of new functionality caused a lot of new bugs in the short term, while the old functionality was fairly stable. This finding is also consistent with other studies of refactoring: bugs often go up in the short term as the new, refactored functionality is being integrated and debugged, but this is not a long-term phenomenon [11]. All the participants confirmed our finding that the ratio of fixed bugs to open bugs was higher after refactoring (in comparison to before). This effect was caused by two related factors: the refactoring both reduced code dependencies and increased productivity. The participants also confirmed our finding that "*the time needed to fix a bug after refactoring is less than the time needed before refactoring*". This was again believe to be the case because the refactoring reduced dependencies, and so bugs were more localized. But the increase in bug-fixing velocity was also aided by the fact that creating the builds became 83.3% faster after refactoring.

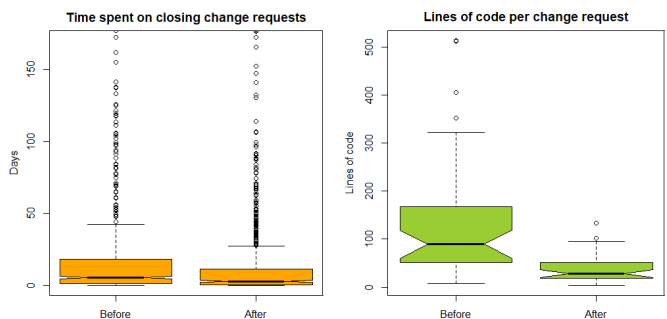

Fig. 4: Analysis of change requests before and after refactoring: Time spent (left) and size of change (right)

When it comes to validating our findings regarding the amount of changes, all participants agreed that "*the number of files changed per bug is significantly less after refactoring*" and "*the number of lines of code to change (churn) per bug is significantly less after refactoring.*" These values were significantly higher before the refactoring and, once again, this was attributed to the higher code complexity and higher inter-file dependencies in the before version.

We wrap up our discussion of our qualitative analysis with a quotation from the Brightsquid's senior architect:

> "*Having an architectural debt analysis report that goes through coupling, circular relationships, and dependencies confirmed our hypotheses and we were able to convey to the top management that we need to do the refactoring as quick as possible.*"

In addition to comparing our results with the perception of developers in previous subsection, we performed semi-structured interviews with two key actors in the company: the product manager and the chief architect. Below we present each of the questions accompanied by the consolidated results of the interview discussions we had with these Brightsquid managers.

*Q: How did you use the report from Phase 2 to decide if/when/how/where to re-factor?*
"We used the report to confirm our own hypotheses on technical debt issues, including circular relationships in our code base.
We did not use specific report findings to determine where we should focus our attention. Instead, there were product and feature changes that drove our decisions regarding code, files and tables to deprecate and/or re-factor."

*Q: To what extend did the report guide you to scrutinize parts or aspects of the architecture that you might not have otherwise focused on?*
The team had a general understanding of the architectural issues inherent in the code base. Business decisions regarding product and feature deprecation drove decisions more than the report.

*Q: Did the results of architectural debt analysis affect your priorities?*
Confirmation and quantification of technical debt through the report made a stronger business case to focus more of the technical teams time on overtly addressing technical debt.

*Q: Did the results of architectural debt analysis affect your*

*refactoring strategies?*

Because of the sheer amount of technical debt, and business decisions made based on feature and product changes, the team determined refactoring strategies without refering to the report.

*Q: Did the report highlight any problems that were not already known/obvious to team members?*

The team was generally aware of the technical debt issues in our code base. The report was very helpful in overtly quantifying the amount and extent of actual technical debt.

*Q: Do you think there is a positive return-on-investment from running architectural debt analysis and the return you received out of it?*

Very much so. Return-on-Investment includes (i) Faster builds: over 50% reduction in building code and (ii) Reduction in files, schemes and tables resulted in simplified architecture. Consequently less time is required to design, write code, test, build and maintain product and features.

*Q: What do you think had the biggest impact from running the debt analysis?*

By confirming both the level and extent of technical debt, it was easier to acquire business commitment to address technical debt through a more cohesive rather than piecemeal approach.

*Q: What would have happened without it?*

In the absence of the report, we would have been more likely to address technical debt in a piecemeal fashion, and dragged this out over a longer period of time.

As a typical software project that has evolved for years, most maintenance costs in the Brightsquid Platform have been focused on just a few file groups, as shown in the Architectural Root analysis. These Roots are typically the root causes of much of the projects technical debt. All after all, as the Brightsquid manager said:

> *"Our code base was a historical record of our quest to find the right market, problem and solution. Architectural debt analysis quantitatively exposed how and where this inflated our cost of change."*

## VI. DISCUSSION

Studying a software organizations efforts for improvement over time allows us to better understand its impact on success [7]. In our longitudinal study, we have covered a period of two years. Any realistic organization does not provide controlled settings, and the same was true for our longitudinal study. A multitude of changes related to resources, development, and business goals occurred concurrently. The impact of such changes is even more substantial for smaller start-up companies such as Brightsquid. Consequently, no strict causality statements can be made. Nevertheless, by triangulating all of the evidence that we collected, a coherent picture emerges. We argue that the architecture debt analysis that we performed had a strong impact on company's decision-makers and this caused them to decide to refactor and guided them in how to focus their refactoring efforts. This refactoring, in turn, helped give the company the ability to mature their technology and better adjust to the competitive health care market.

Refactoring and removing technical debt are a priority and an ongoing effort for Brightsquid, but these needed to be balanced against the business's other priorities—delivering customer-facing features. The Brightsquid stakeholders have seen, however, the significant benefits to the business and the development team in reducing complexity, removing architectural flaws, and simplifying or removing obsolete business rules. The refactoring efforts helped to increase Brightsquid team's productivity and ability to adapt quickly to new business requirements.

We argue that the overall message presented in this paper—that architecture debt was weighing Brightsquid down and that the refactoring removed substantial portions of this debt—is the result of combining various independent streams of reasoning and evidence:

1) **Quantitative analysis:** This analysis revealed improvements to both the architecture and the maintainability of the platform software. This analysis was based on measuring key attributes of the system's software architecture—its complexity and architecture flaws and roots—and key outcome measures, such as lines of code, bugs, and velocity. Our study was limited to Brightsquid's platform project and focused on its Java code, but this was justifiable as the platform is the most critical software component for the company. Still, a more comprehensive analysis including other parts of the system, is needed to increase the validity of our findings.

2) **Qualitative analysis:** This evidence was based on performing a series of interviews with key members of the project team, asking for their perceptions surrounding the relevance, utility, and accuracy of the architectural debt analysis. This allowed us to assess the perceived value of the analyses that we performed. Even though we had just four participants, the clear trend was that the architectural analysis confirmed our hypotheses and gave Brightsquid the evidence that they needed for their decision making. In addition, the analysis allowed the team to argue for the urgency of performing a substantial refactoring. While these forms of interviews are subject to validity threats, when combined with the quantitative results a consistent picture emerges.

3) **Preponderance of evidence:** Finally, we argue that the results achieved here, in terms of the reduction in architectural flaws and the subsequent gains in productivity, are consistent with a large and growing body of research evidence—that technical and architectural debt matters and paying it down can catalyze substantial productivity improvements (e.g. [2], [3], [5], [16], [21]). The value and insight gained from running architectural debt analyses has been shown in other studies. In particular it has been shown that the number of architectural flaws per file is very strongly correlated with bugs, changes, and churn/effort [6], [16]. In this study, for the first time, we performed multiple analyses of a *single* system over time. In particular, we measure the architectural debt and productivty measures before and after refactoring.

Thus while we can not argue for strict causality—that the refactoring *caused* the productivity improvements—all of the available evidence—qualitative, quantitative, and our prior research corpus of results—points in the same direction. This gives us confidence in arguing that reducing architectural debt has an influence on productivity and quality measures that projects care about.

Having seen the value of this analysis, Brightsquid is now considering embarking on a program of *continuous* architecture measurement and benchmarking. Measurement, particularly if it is fully automated and connected with project triggers (for example, a nightly build), can quickly notify project stakeholders if there has been a degradation in the architecture—perhaps the introduction of a new architecture flaw. Benchmarking can help a company understand how it is situated in terms of its competitive landscape. So, for example, a company can track its DL score over time and use this to determine if any mitigations are required. This is no different than what you do when you go to the doctor. The doctor analyzes your health via a spectrum of analyses and tests and notes differences from your last checkup (for example, a dramatic increase in blood pressure or cholesterol levels) and uses broad benchmarks to decide if an intervention is necessary (for example, a total cholesterol level of 220 is slightly high but within the normal range for adults). All of the architectural analyses presented in this paper are fully automated and so they can provide exactly the data needed for such measurement and benchmarking (e.g. [17]).

## VII. Related work

In this section, we consider previous work related to our case study.

*Architectural Debt Analysis*. In the past decade, a number of methods have been proposed to analyze technical debt [4], [26], [27] within software systems. DV8 and Titan have been used to analyze and detect architectural debt in both open source and (closed source) industrial projects. Kazman et. al. [10] presented the experiences of detecting and quanti- fying architectural debts in Softserve, as well as a return on investment estimation of the benefit that would accrue to potential refactoring activities. The recent work of Mo et al [18] reported the experience of applying the DV8 tool suite to eight projects of various sizes and domains within ABB Corporation, and showed that the architecture analysis helped practitioners to make decisions on if and where to refactor.

Carriere et. al. [3] also proposed and applied a cost- benefit model to estimate benefits of reducing the level of coupling in an e-conmerce architecture. Their work focused on coupling only, rather than on identifying architecture flaws and pinpointing the locations to refactor.

Curtis et. al. [5] proposed a model to estimate the amount technical debt in terms of cost calculated from source code static relations. Nord et. al. [21] created a formula to estimate the impact of technical debt on architecture, which could be used to optimize long-term product evolution. Similarly to Curtis et. al., their work does not detect the location and specific problems that need to be treated.

Martini and Bosch [15] proposed and validated AnaCon-Debt, a method that has been applied to 12 case systems within 6 companies. Their experience showed that it can help practitioners to decide if and when to refactor architectural debt items. Different from Titan and DV8, the architectural debt in their work had to be manually identified beforehand by architects.

Of all these previous cases studies, none of them presented the experiences before and after the actual refactoring activities, although many of them demonstrated that their approaches can help the team to make refactoring decisions. The work we are presenting here is, to our knowledge, the first report of analyzing a system before and after refactoring, in terms of architecture debt variations and maintenance effort variations, both quantitatively and qualitatively.

## VIII. Summary and Conclusions

As described by Runeson et al. [22], case studies in software engineering are often just examining small-scale systems with relatively low research effort. There is a lack of longitudinal case study research (with some notable exceptions such as [12]) which are particularly appropriate for larger scale, complex phenomena where there is a need to collect both quantitative and qualitative data. As mentioned by [22], "software engineering research community both recognizes the demands of longitudinal case study research but also that the community can only rarely allocate sufficient resources toward such studies".

Our current study covers a period of two years and reports on the process and major findings from identifying and paying down architectural debt. In this study we have observed substantial improvements made over the period of intervention and analysis. Understanding the architectural flaws and initiating refactoring helped the company to reduce their build time by about 83%, reduced the average time to resolve issues by 72%, and reduce bug-fixing effort from an average of 102 LOC per bug to just 34. How do we know that these improvements all accrued to the architectural debt repayment? We can not claim causality, of course, based on a single case study. However, we can make an argument based on a preponderance of evidence all pointing in the same direction. The number and size of the architecture flaws went down, productivity measures went up, the key stakeholders felt that the refactoring benefited their ability to manage the code base. And, finally, all of this is consistent with the evidence collected over the analyses of hundreds of open source and industrial projects in our prior work which shows a very strong correlation between architectural flaws and productivity measures [6], [16]. Thus, although this is just a single case study, it presents a consistent picture—that architectural debt matters and that it is possible to pay it down, via refactoring, and achieve significant benefits as a result.

Many start-ups embrace agile software development methodologies, where the typical attitude is that solution architecture evolves organically. Evolution is at the mercy a continually changing environment, so key start-up characteristics comprise survival, flexibility, speed, and a revolving door of opportunities, stakeholders and employees. The research presented here is part of a long term project with BrightSquid to improve their software processes. For example we developed ESSMArT for automatic escalation and summarization of user requests [19] and implemented a tagging method for better estimation of effort needed to close a change request [20]. The research reported here is a second major step towards our goal of *Intelligent Change Management*.

The job of the start-up is to find a sustainable business model—in other words, to discover an important and urgent problem, that a defined and accessible segment of customers will pay for. But this often seems to be inherently in conflict with a disciplined approach towards software engineering in general and with the management of technical debt in particular. The final outcome of this study is that it made believers out of Brightsquid's management. Thus we will conclude with a quotation from the product owner.

"The approach expressed in the paper to analyze architectural debt provides immutable guide-stones to navigate through the mutable destinations on the road to discovering the actual sustainable business model. If we declare the components of your analysis to be essential non-functional requirements (clique, package cycle, improper inheritance, modularity violation, crossing, unstable interface) with overtly declared targets, then regardless of the markets, problems or solutions visited on the journey, the disciplined start-up that regularly manages these non-functional requirements is much better equipped to avoid architectural debt accumulation in the first place. As long as acceptable thresholds are maintained, the start-up can be confident that the architecture supports, rather than thwarts success. Quick and dirty fixes can consequently be exposed as what they truly are - incremental, short-sighted steps down a slippery slope that inflates the costs of change. In short, *An ounce of prevention is worth a pound of cure.*"

## REFERENCES

[1] A. Begel and T. Zimmermann. Analyze this! 145 questions for data scientists in software engineering. In *Proceedings of the 36th International Conference on Software Engineering*, pages 12–23. ACM, 2014.

[2] N. Brown, Y. Cai, R. Kazman, M. Kim, P. Kruchten, E. Lim, A. MacCormack, R. Nord, I. Ozkaya, R. Sangwan, C. Seaman, K. Sullivan, and N. Zazworka. Managing technical debt in software-reliant systems. In *FSE/SDP Workshop on the Future of Software Engineering Research at ACM SIGSOFT FSE-18*, 2010.

[3] J. Carriere, R. Kazman, and I. Ozkaya. A cost-benefit framework for making architectural decisions in a business context. In *Proc. 32nd International Conference on Software Engineering*, pages 149–157, 2010.

[4] W. Cunningham. The WyCash portfolio management system. In *Addendum to Proc. 7th*, pages 29–30, Oct. 1992.

[5] B. Curtis, J. Sappidi, and A. Szynkarski. Estimating the principal of an application's technical debt. *IEEE Software*, 29(6):34–42, 2012.

[6] Q. Feng, R. Kazman, Y. Cai, R. Mo, and L. Xiao. An architecture-centric approach to security analysis. In *Proc. 13th Working IEEE/IFIP International Conference on Software Architecture*. IEEE, 2016.

[7] B. Fitzgerald and T. O'Kane. A longitudinal study of software process improvement. *IEEE software*, 16(3):37–45, 1999.

[8] S. J. Kabeer, M. Nayebi, G. Ruhe, C. Carlson, and F. Chew. Predicting the vector impact of change-an industrial case study at brightsquid. In *Empirical Software Engineering and Measurement (ESEM), 2017 ACM/IEEE International Symposium on*, pages 131–140. IEEE, 2017.

[9] R. Kazman, Y. Cai, R. Mo, Q. Feng, L. Xiao, S. Haziyev, V. Fedak, and A. Shapochka. A case study in locating the architectural roots of technical debt. In *Software Engineering (ICSE), 2015 IEEE/ACM 37th IEEE International Conference on*, volume 2, pages 179–188. IEEE, 2015.

[10] R. Kazman, Y. Cai, R. Mo, Q. Feng, L. Xiao, S. Haziyev, V. Fedak, and A. Shapochka. A case study in locating the architectural roots of technical debt. In *Proc. 37th International Conference on Software Engineering*, May 2015.

[11] M. Kim, D. Cai, and S. Kim. An empirical investigation into the role of api-level refactorings during software evolution. In *Proc. 33rd International Conference on Software Engineering*, May 2011.

[12] J. Li, N. B. Moe, and T. Dybå. Transition from a plan-driven process to scrum: a longitudinal case study on software quality. In *Proceedings of the 2010 ACM-IEEE international symposium on empirical software engineering and measurement*, page 13. ACM, 2010.

[13] A. MacCormack, J. Rusnak, and C. Baldwin. Exploring the duality between product and organizational architecture: A test of the mirroring hypothesis. Working Paper 08-039, Harvard Business School, Oct. 2008. http://www.hbs.edu/research/pdf/08-039.pdf.

[14] A. MacCormack, J. Rusnak, and C. Y. Baldwin. Exploring the structure of complex software designs: An empirical study of open source and proprietary code. *Management Science*, 52(7):1015–1030, July 2006.

[15] A. Martini and J. Bosch. An empirically developed method to aid decisions on architectural technical debt refactoring: Anacondebt. In *Proc. 38th International Conference on Software Engineering*, May 2016.

[16] R. Mo, Y. Cai, R. Kazman, and L. Xiao. Hotspot patterns: The formal definition and automatic detection of architecture smells. In *Proc. 12th Working IEEE/IFIP International Conference on Software Architecture*, May 2015.

[17] R. Mo, Y. Cai, R. Kazman, L. Xiao, and Q. Feng. Decoupling level: A new metric for architectural maintenance complexity. In *Proc. 38thInternational Conference on Software Engineering*, pages 499–510, 2016.

[18] R. Mo, W. Snipes, Y. Cai, S. Ramaswamy, R. Kazman, and M. Naedele. Experiences applying automated architecture analysis tool suites. In *Proc. 33th IEEE/ACM International Conference on Automated Software Engineering*, pages 197–208, Sept. 2018.

[19] M. Nayebi, L. Dicke, R. Ittyipe, C. Carlson, and G. Ruhe. Essmart way to manage user requests. *arXiv preprint arXiv:1808.03796*, 2018.

[20] M. Nayebi, S. J. Kabeer, G. Ruhe, C. Carlson, and F. Chew. Hybrid labels are the new measure! *IEEE Software*, 35(1):54–57, 2018.

[21] R. L. Nord, I. Ozkaya, P. Kruchten, and M. Gonzalez-Rojas. In search of a metric for managing architectural technical debt. In *2012 Joint Working IEEE/IFIP Conference on Software Architecture and European Conference on Software Architecture*, pages 91–100, 2012.

[22] P. Runeson, M. Host, A. Rainer, and B. Regnell. *Case study research in software engineering: Guidelines and examples*. John Wiley & Sons, 2012.

[23] L. Xiao, Y. Cai, and R. Kazman. Design rule spaces: A new form of architecture insight. In *Proc. 36rd International Conference on Software Engineering*, 2014.

[24] L. Xiao, Y. Cai, and R. Kazman. Titan: A toolset that connects software architecture with quality analysis. In *22nd*, 2014.

[25] L. Xiao, Y. Cai, R. Kazman, R. Mo, and Q. Feng. Identifying and quantifying architectural debt. In *Proceedings of the 38th International Conference on Software Engineering*, pages 488–498. ACM, 2016.

[26] L. Xiao, Y. Cai, R. Kazman, R. Mo, and Q. Feng. Identifying and quantifying architectural debt. In *Proc. 38thInternational Conference on Software Engineering*, pages 488–498, 2016.

[27] N. Zazworka, A. Vetro, C. Izurieta, S. Wong, Y. Cai, C. Seaman, and F. Shull. Comparing four approaches for technical debt identification. pages 1–24, 2013.

[28] M. Nayebi, et al. "Analytics for Software Project Management--Where are We and Where do We Go?." Automated Software Engineering Workshop (ASEW), IEEE, 2015.